\documentclass[aps,prb,amsmath,twocolumn,superscriptaddress]{revtex4-2}

\usepackage{amsmath}
\usepackage{amssymb}
\usepackage{hyperref}
\usepackage{tikz}

\usepackage{graphicx}
\usepackage{bm} 
\usepackage{url}
\usepackage{verbatim}   

\usepackage{color}

\usepackage[percent]{overpic}
\usepackage[export]{adjustbox}
\usepackage{stackengine}
\usepackage{appendix}

\usepackage{placeins} 
\usepackage{flafter}  

\begin{document}

\title{Relaxation  of  maximally entangled quantum states of two nonequivalent nuclear spins in a liquid}

\author{Georgiy Baroncha}
\email{georgiy.baroncha@uni-leipzig.de}
\affiliation{Felix Bloch Institute for Solid State Physics, University of Leipzig,
Linnéstraße 5, 04103 Leipzig, Germany}
\affiliation{Terra Quantum AG, Kornhausstrasse 25, 9000 St. Gallen, Switzerland}
 
\author{Alexander Perepukhov}
\affiliation{Terra Quantum AG, Kornhausstrasse 25, 9000 St. Gallen, Switzerland}
 
\author{Boris V. Fine}
\email{boris.fine@uni-leipzig.de}
\affiliation{Institute for Theoretical Physics, University of Leipzig,
Br{\"{u}}derstr. 16, 04103 Leipzig, Germany}%

\date{\today}

\begin{abstract}
We investigate both experimentally and theoretically the relaxation of pseudo-pure maximally entangled states (Bell states) of two nuclear spins $^1$H~--$^{\ 13}$C belonging to a molecule in a liquid.  The Bell states are obtained by a method based on a detuned Hartmann-Hahn cross-polarization condition. Their entangled character is verified by quantum-state tomography. Our relaxation measurements reveal different relaxation rates for different Bell states. We interpret this difference as originating from cross-correlations between different relaxation mechanisms, thereby demonstrating that the measurements of the differential relaxation of Bell states are potentially useful for advanced NMR characterization of liquids.

\end{abstract}

\maketitle

\section{Introduction}

Maximally entangled states of two quantum spins 1/2, known as ``Bell states", have been of interest for nuclear magnetic resonance (NMR)  for decades. These states are often exploited in the context of the quantum computing agenda\cite{CORY199882, Chuang1998}, but their utility in NMR is much broader, as they can be used for advanced preparation and characterization of nuclear spin systems. Among the Bell states,  the best known is, perhaps, the singlet state.   It was, in particular, shown both experimentally \cite{doi:10.1021/ja0490931, doi:10.1073/pnas.1010570107, doi:10.1126/science.1167693,  article1, CLAYTOR201481} and theoretically \cite{doi:10.1063/1.1893983,   doi:10.1063/1.2778429, doi:10.1063/1.5074199} that the nuclear singlet relaxation time can be an order of magnitude longer than the relaxation time of a single spin. Singlets can also be utilized as a part of the NMR hyperpolarization routine\cite{soton347308},  in algorithmic cooling of nuclear spins \cite{doi:10.1063/5.0006742}, in the measurement of diffusion rate by NMR spectroscopy \cite{https://doi.org/10.1002/cmr.a.20100} or for the study of very slow dynamic processes \cite{doi:10.1021/ja0647396}. Singlet-filtered spectroscopy is another promising application \cite{doi:10.1126/sciadv.aaz1955} in NMR studies.  

Given the above successes, the NMR investigations of Bell states have been focused so far largely on the singlets of nuclear spins with the same gyromagnetic ratios. 
In this article, we present a comparative relaxation study of a singlet state and other Bell states for a pair of nuclear spins having different gyromagnetic ratios in a strong magnetic field.
The relaxation of spin singlets in a similar setting with a focus on the low magnetic fields was previously investigated in \cite{PhysRev.128.2042}.  Other Bell states of a nuclear spin pair have been investigated in Ref.\cite{Samal_2010} but not in the direction of comparative relaxation study. 

In strong magnetic fields, the relaxation of Bell states of two different nuclear spin species is not expected to be much longer than the usual $T_1$-relaxation of each of the two spins\cite{PhysRevLett.112.077601}. Yet these two relaxations can be measurably different, and this difference can depend on the properties of the microscopic environment of the spin pair that would not be accessible by the standard $T_1$ measurements. Our motivation was to examine experimentally, whether the measurable differences exist, and whether they can be understood theoretically. In this article, we arrive at the positive answers to the both questions. These answers affirm the analysis of Refs.\cite{KUMAR2000191,GHOSH2005125}. Along the way, we also introduce a method for generating pseudo-pure Bell states based on detuned Hartmann-Hahn double resonance conditions.  

Below in Section~\ref{theoretical}, we describe the theoretical setting and the theoretical basis of our method for generating the Bell states.  In Section~\ref{exp}, we introduce our experimental setup and the experimental pulse sequences.  The results of our relaxation measurements are presented and theoretically interpreted in Section~\ref{relaxation}.  Section~\ref{conclusions} contains the conclusions.


\section{Theoretical description}
\label{theoretical}
\subsection{General formulation}
We consider two nuclear spins $1/2$,  $\mathbf{S}_1$ and $\mathbf{S}_2$,  with different gyromagnetic ratios $\gamma_1$ and $\gamma_2$  belonging to a molecule in a liquid placed in a strong static magnetic field $B$ directed along the $z$-axis. The two spins interact {\it via}  the $J$-coupling with constant $J$. The secular part of the spin pair Hamiltonian in the laboratory reference frame is thus
\begin{equation}
\label{1}
    \mathcal{H}_{\text{lab}} = - \Omega_1 S_{1z} - \Omega_2 S_{2z} + J {S}_{1z} {S}_{2z},
\end{equation}
where $\Omega_1 = \gamma_1 B$ and $\Omega_2 = \gamma_2 B$. We set $\hbar = 1$.

In what follows, we use the ``double-rotating" reference frame, which means that the spin projections (${S}_{1x}$, ${S}_{1y}$) and (${S}_{2x}$, ${S}_{2y}$) are defined in the reference frames rotated around the $z$ axis with the respective frequencies $-\Omega_1$ or $-\Omega_2$. The above transformation into the double-rotating frame converts the Hamiltonian (\ref{1}) into $\mathcal{H}_0 =  J {S}_{1z} {S}_{2z}$.

The initial equilibrium density matrix of the spin pair in both the laboratory and the double-rotating frames is
\begin{equation}
\label{rho-eq}
    \rho_{eq} =  \frac{1}{4} \mathbb{I} + \varepsilon_1 S_{1z} + \varepsilon_2 S_{2z}  ,
\end{equation}
where $\mathbb{I}$ is a $4 \times 4$ identity operator,  $\varepsilon_1 = \Omega_1/(4 k_B T)  $,  \mbox{$\varepsilon_2 = \Omega_2/(4 k_B T) $}, with $k_B$ and $T$ being respectively the Boltzmann constant and the temperature. Eq.(\ref{rho-eq}) implies the high-temperature approximation.
Since the term $\mathbb{I}/4$ in the density matrix remains invariant under time evolution and does not contribute to the NMR signal, it is omitted in all formulas for the density matrices in the rest of this article. 

The spin pair is defined to be in a  pseudo-pure state (PPS) when the correction to the above identity term is proportional to $|\Psi \rangle \langle\Psi | $, where $|\Psi \rangle$ is a wave function of some pure state of the pair. The equilibrium state (\ref{rho-eq}) is not a PPS but rather a mixed state. The goal of this work is to create an initial pseudo-pure Bell state, i.e. a state, where $|\Psi \rangle$ is a maximally entangled wave function, and then investigate the relaxation of such a state.


\subsection{Preparation of Bell states using detuned Hartmann-Hahn double resonance condition}
    \label{theorynon}

As the quantum basis of the spin pair, we use one singlet and three triplet states (defined in the double rotating frame):  
\begin{equation}
    \begin{split}
    \label{2}
        |T_{+1,x}\rangle = |\uparrow_x \uparrow_x \, \rangle, \qquad \qquad \quad \\
        |T_{0,x}\rangle = \frac{1}{\sqrt{2}}(|\uparrow_x \downarrow_x \,\rangle  + |\downarrow_x \uparrow_x \,\rangle),\\
         |S_{0}\rangle =\frac{1}{\sqrt{2}}(|\uparrow_x \downarrow_x \,\rangle  - |\downarrow_x \uparrow_x \,\rangle),\\
        |T_{-1,x}\rangle = |\downarrow_x \downarrow_x \,\rangle. \qquad \qquad \quad \\
    \end{split}
\end{equation} 
where spin projections are quantized along the $x$-axis (the direction of the applied rf-fields below) , $|T_{s,x}\rangle$ are the symmetric triplet states and $|S_{0}\rangle$ is the antisymmetric singlet state.

There exist four mutually orthogonal Bell states of the pair, namely,  $|S_{0}\rangle$, $|T_{0,x}\rangle$, and also  
\begin{equation}
    \begin{split}
    \label{2.1} 
        \vert \psi_{+,x}\rangle \equiv \frac{1}{\sqrt{2}} (\vert \uparrow_x \uparrow_x \, \rangle +\vert \downarrow_x \downarrow_x \, \rangle), \\
        \vert \psi_{-,x}\rangle \equiv \frac{1}{\sqrt{2}} (\vert \uparrow_x \uparrow_x \, \rangle - \vert \downarrow_x \downarrow_x \, \rangle). \\ 
        \end{split}
\end{equation} 
In the experimental part, we will eventually rotate the quantization axis from $x$ to $z$ and deal with the analogously defined Bell states $\vert S_0 \, \rangle$,  $\vert T_{0,z} \, \rangle$, $\vert \psi_{-,z} \, \rangle$, and  $\vert \psi_{+,z} \, \rangle$.

In order to experimentally generate the entangled PPSs, we start with the non-entangled PPSs 
$\vert \downarrow_x \uparrow_x \, \rangle $, 
$\vert \uparrow_x \downarrow_x \, \rangle $, 
$\vert \uparrow_x \uparrow_x \, \rangle $, and
$\vert \downarrow_x \downarrow_x \, \rangle $, which are to be obtained by the technique of Ref.\cite{CORY199882}.

Our procedure for converting non-entangled PPSs into entangled ones is based on the detuned Hartmann-Hahn (DHH) double resonant condition. The procedure requires the application of two resonant rf-fields at frequencies $\Omega_1 $ and $\Omega_2$ with amplitudes $h_1$ and $h_2$ respectively.  
In the presence of these rf fields, the Hamiltonian in the double-rotating frame becomes \begin{equation}
    \label{3}
    \mathcal{H}  = -\omega_1  S_{1x} - \omega_2  S_{2x} + JS_{1z}S_{2z},
\end{equation}
where $\omega_1  = \gamma_1 h_1 $ and $\omega_2  = \gamma_2 h_2$. 

The Hilbert space of two spins 1/2 can be decomposed into the so-called ``zero quantum space" with the basis $\vert \downarrow_x \uparrow_x \, \rangle $ and 
$\vert \uparrow_x \downarrow_x \, \rangle $, and the ``double quantum space" with the basis $\vert \uparrow_x \uparrow_x \, \rangle $ and
$\vert \downarrow_x \downarrow_x \, \rangle $. The Hamiltonian~(\ref{3}) does not mix the zero and the double spaces. 

The dynamics in the zero quantum space is controlled by parameter $\Delta \equiv \omega_1 - \omega_2$, while the dynamics in the double quantum space is controlled by $\Sigma \equiv \omega_1 + \omega_2$. The entangled states $|S_{0}\rangle$ and $|T_{0,x}\rangle$ belong to the zero quantum space. In order to obtain them, one needs the Hamiltonian ~(\ref{3}) with $\Delta  = J/2$ to act on, respectively,
$\vert \downarrow_x \uparrow_x \, \rangle $ and 
$\vert \uparrow_x \downarrow_x \, \rangle $ during time $t = \frac{\pi\sqrt{2} }{J }$. The value of $\Sigma$ does not matter here.
On the other hand, the states $\vert \psi_{+,x}\rangle$ and  $\vert \psi_{-,x}\rangle$ belong to the double quantum space. They can be obtained once the Hamiltonian ~(\ref{3}) with ${\Sigma  = J/2}$ acts, respectively, on $\vert \uparrow_x \uparrow_x \, \rangle $ and
$\vert \downarrow_x \downarrow_x \, \rangle $ 
during time ${t = \frac{\pi \sqrt{2} }{J}}$. The value of $\Delta$ is arbitrary in this case.

Let us remark here that the detuned Hartmann-Hahn condition does not require $\omega_1, \omega_2 \gg J$. On the contrary, the condition $\Sigma  = J/2$ in the double quantum space implies that $\omega_1, \omega_2 \sim J$. The condition $\Delta  = J/2$ can be satisfied with both $\omega_1, \omega_2 \sim J$ and $\omega_1, \omega_2 \gg J$. 

\begin{figure}[h!]
   
    \centering
    \begin{tikzpicture}
    \node[inner sep=0pt] (duck) at (0,0)
    {\includegraphics[width=0.85\columnwidth]{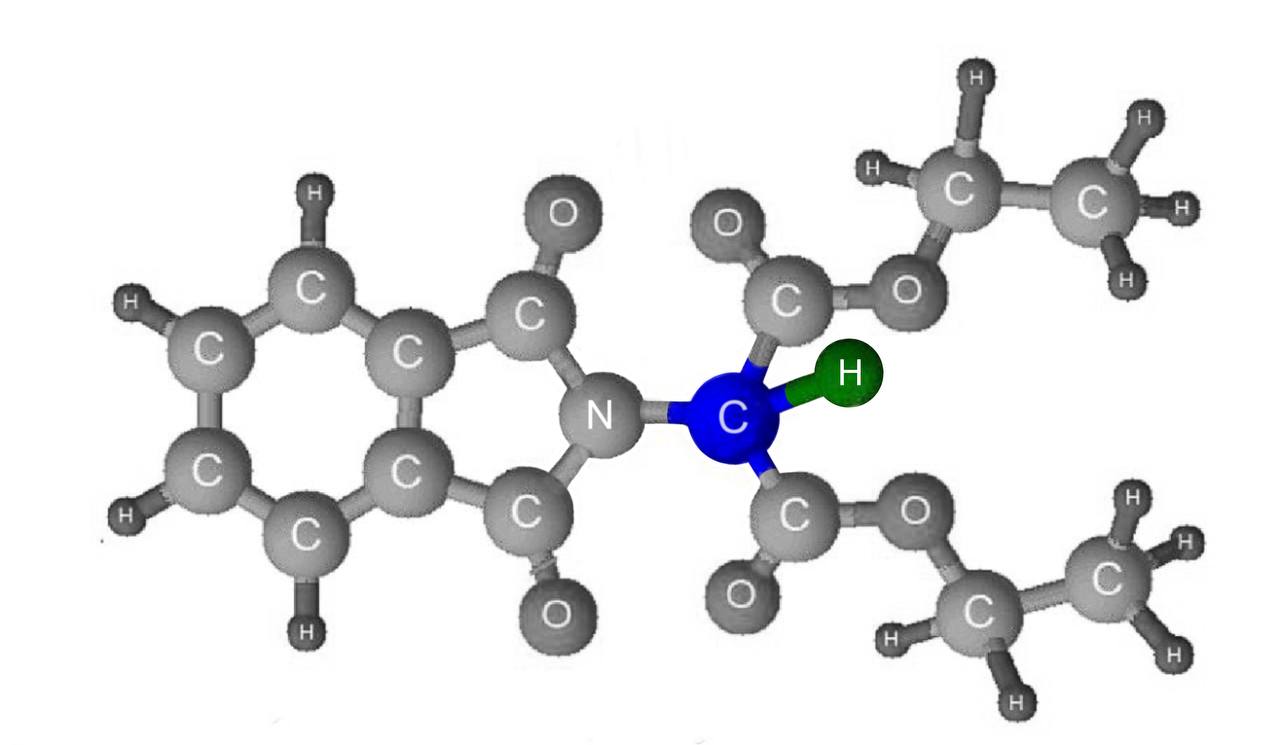}};
    \node[align=center,fill=white] at (-3.8, 1.3) {\textbf{(a)}};
    \draw[blue] (0.55,-0.25) circle (10pt);
     \draw[green] (1.21,-0.00) circle (7pt);
    \end{tikzpicture}
    
    \begin{tikzpicture}
    \node[inner sep=0pt] (duck) at (0,0)
    {\includegraphics[width=0.85\columnwidth]{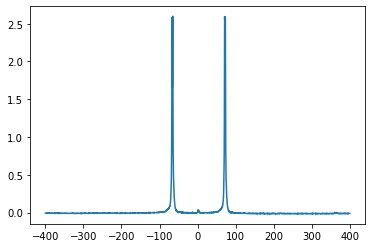}};
    \node[align=center,fill=white] at (-3.8, 2.5) {\textbf{(b)}};
    \node[align=center,fill=white] at (-3.9, 0.3) {{\rotatebox {90}{\text{$g(\omega)$ [a.u.]} }}};
    \end{tikzpicture}
    \text{\qquad \quad $\omega/2\pi$ [Hz] }
    \begin{tikzpicture}
    \node[inner sep=0pt] (duck) at (0,0)
    {\includegraphics[width=0.85\columnwidth]{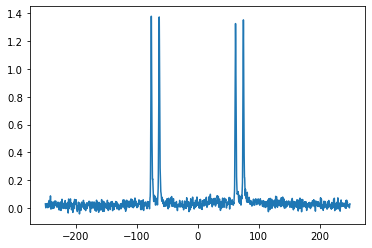}};
    \node[align=center,fill=white] at (-3.8, 2.5) {\textbf{(c)}};
    \node[align=center,fill=white] at (-3.9, 0.3) {{\rotatebox {90}{\text{$g(\omega)$ [a.u.]} }}};
    \end{tikzpicture}
    \text{\qquad \quad $\omega/2\pi$ [Hz] }
    \caption{
\label{fig1}
 (a)  Diethylphthalimidomalonate($-2-^{13}$C,$^{15}$N molecule with the measured pair of nuclear spins $^1$H~--$^{\ 13}$C highlighted by circles.  (b) Spectrum $g(\omega)$ of the circled $^1$H, where $\omega$ is the frequency offset.   (c) Spectrum $g(\omega)$  of the circled $^{13}$C. The small splitting of each of the two peaks is due to the $J$-coupling with the nearby $^{15}$N spin, which is neglected in our analysis.
}
\end{figure}
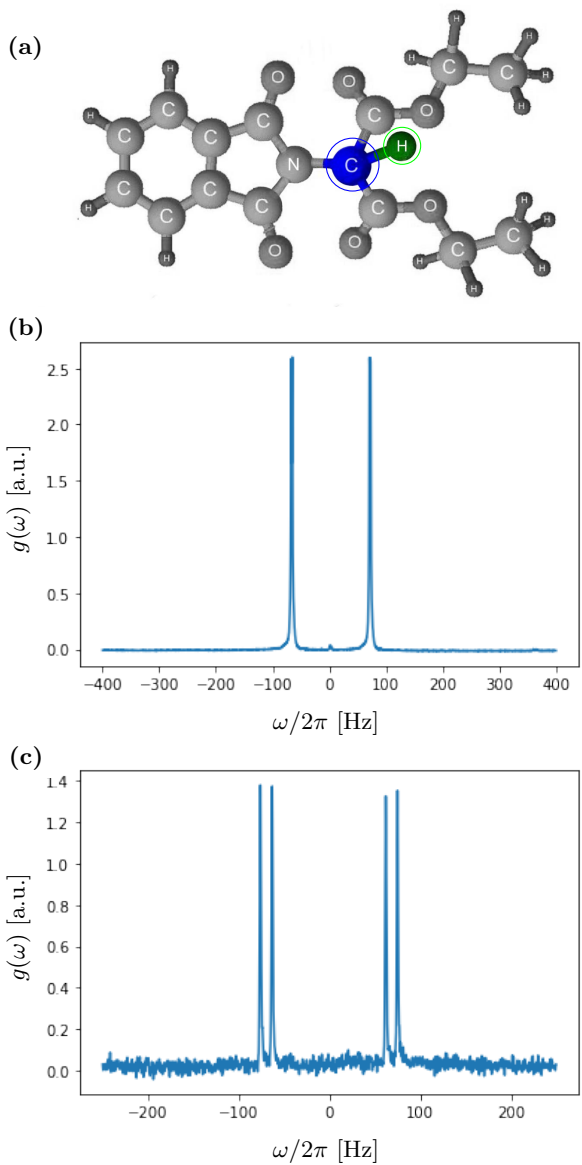

 \section{Experimental setting }
 \label{exp}
 
We measure the pair of adjacent nuclear spins 1/2, $^1$H and $^{13}$C belonging to the molecule diethylphthalimidomalonate($-2-^{13}$C,$^{15}$N) in the solution C$_6$D$_6$. The molecular structure is shown in Fig.~\ref{fig1}(a) with the selected $^1$H~--$^{\ 13}$C pair indicated. The molecule is 99.5 percent enriched by $^{13}$C and $^{15}$N at the indicated positions. The $J$-coupling constant for the $^1$H~--$^{\ 13}$C pair is $J/(2 \pi) = 138$~Hz.
The $J$-couplings of  $^{13}$C and $^1$H  to $^{15}$N are much smaller: $13.1$~Hz and $1.7$~Hz, respectively. The $J$-coupling of the $^1$H~--$^{\ 13}$C pair to other nuclear spins belonging to the molecule is even weaker, as it was not resolved in the measured spectra. It is thus justifiable to treat the $^1$H~--$^{\ 13}$C pair as isolated on the time scale of the order of $1/J$ required to prepare and characterize the Bell states. 
The molecule and the equilibrium NMR spectra of $^1$H and $^{13}$C are shown on  \ref{fig1}.

To connect to the theoretical  Section~\ref{theoretical}, the variables  $\mathbf{S}_1$, $\gamma_1$ and $\omega_1$ refer to  $^1$H, while  $\mathbf{S}_2$, $\gamma_2$ and $\omega_2$ refer to  $^{13}$C.

The NMR experiments were performed using NMR spectrometer Varian Inova 500 with the magnetic field $B  = 11.7 $ T corresponding to the $^1$H Larmor frequency $\Omega_1 = 500 $~MHz. 

 \section{Preparation and characterization of Bell states}
\label{preparation}
 
 \subsection{Preparation pulse sequences}
 \label{Pulse sequence}
  
\begin{figure*} 
    \centering
    \begin{tikzpicture}
    \node[inner sep=0pt] (duck) at (0,0)
    {\includegraphics[width=1.8\columnwidth]{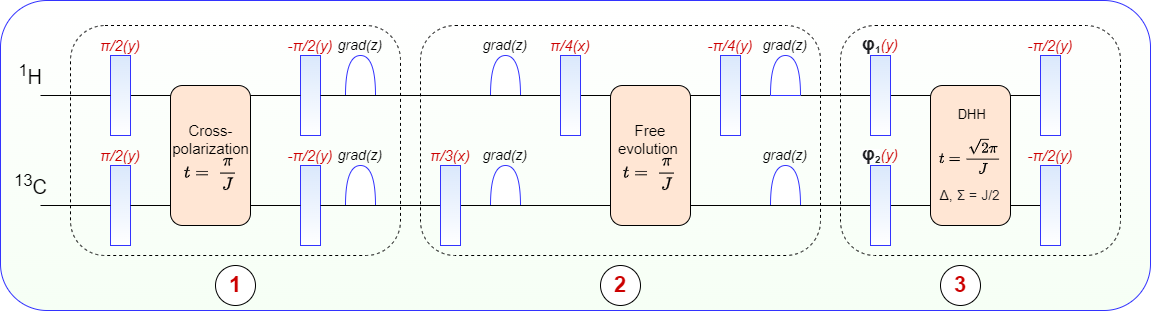}}; 
    \end{tikzpicture}

    \caption{
\label{fig2}
Experimental pulse sequence described in the text: steps 1, 2, and 3 are indicated by the respective circled numbers.   CP$[t=\pi /J]$ is the Hartmann-Hahn resonant cross-polarization during the time $t=\pi  /J$.  DHH$[t=\pi \sqrt{2} /J]$ is the evolution during time $t=\pi \sqrt{2} /J$  under a detuned Hartmann-Hahn condition with either  $\Delta = J/2$ or $\Sigma = J/2$ as appropriate for the target Bell state  as indicated in Table~\ref{table1}. The pulses $\varphi_1$ and  $\varphi_2$ are also listed in the first column of Table~\ref{table1}.
}
\end{figure*}

Our experimental procedure for preparing pseudo-pure Bell states starts with the equilibrium density matrix (\ref{rho-eq}) and then proceeds as follows 
 \begin{equation} \begin{split}
    \varepsilon_1 S_{1z} + \varepsilon_2 S_{2z}  \  \xrightarrow {\quad \text{Step 1} \quad}   \ \frac{\varepsilon_1 + \varepsilon_2}{2} (S_{1z} + S_{2z})
\label{17}
    \\ 
    \xrightarrow {\quad \text{Step 2} \quad} 
    \frac{\varepsilon_1 + \varepsilon_2}{2}
    |\uparrow_z \uparrow_z \, \rangle \langle \, \uparrow_z \uparrow_z  \! | \xrightarrow {\quad \text{Step 3} \quad} \mbox{\  Bell state},
    \end{split} 
\end{equation}
 where the formulas before and between the arrows represent the density matrices with the trivial part $\frac{1}{4} \mathbb{I}$ dropped.

The goal of step 1  is to equalize the spin polarizations of $^1$H and $^{13}$C thereby preparing the system for step 2. Here, in order to minimize the loss of the signal, we do not use the simplest routine, namely, the partial rotation of $^1$H spin followed by a gradient pulse on $^1 H$, but rather we first apply $(\pi/2)_{1y}$,$(\pi/2)_{2y}$ pulses followed by the resonant Hartmann-Hahn pulse with $\omega_1 = \omega_2 = \frac{\sqrt{15}}{4} J$ during time $\pi/J$ to partially switch the polarizations of $^1$H and $^{13}$C, then apply $(-\pi/2)_{1y}$,$(-\pi/2)_{2y}$, and then apply the gradient pulse. 

Step 2 employs the sequence proposed in Ref.~\cite{CORY199882}
that starts with an equally polarized mixed state and then generates a pseudo-pure non-entangled state ${|\uparrow_z \uparrow_z \, \rangle \langle \, \uparrow_z \uparrow_z  \! | }$.

In step 3, we rotate the state $|\uparrow_z \uparrow_z \, \rangle  $ to an initial non-entangled state indicated in Table~\ref{table1}  and then apply the DHH pulses defined in Section~\ref{theorynon} to generate the desired Bell states. The specific DHH parameters for the Bell states $\vert S_0 \, \rangle$,  $\vert T_{0,x} \, \rangle$, $\vert \psi_{-,x} \, \rangle$, and  $\vert \psi_{+,x} \, \rangle$ are given  in Table~\ref{table1}.

Finally, we apply two
$ ( -\pi/2)_{1y}$, $ ( -\pi/2)_{2y}$ pulses, thereby rotating the states   $\vert T_{0,x} \, \rangle$, $\vert \psi_{-,x} \, \rangle$, and  $\vert \psi_{+,x} \, \rangle$ into, respectively,
$\vert T_{0,z} \, \rangle$, $\vert \psi_{-,z} \, \rangle$, and  $\vert \psi_{+,z} \, \rangle$. No rotation was necessary for the  state $\vert S_0 \, \rangle$ due to its isotropic character.


\begin{table}[h!] 
\begin{tabular}{ c|c|c|c} 
\hline
Pulse parameters & Pseudo-pure states  & DHH & Resulting\\
$\varphi_1$ ,  $\varphi_2$ & after $\varphi_1, \varphi_2$  pulses & parameters & Bell state\\
\hline
$ (-\pi/2)_{1y}$, $ (-\pi/2)_{2y}$  & $\vert \downarrow_x \downarrow_x \, \rangle $  & $\Sigma = J/2$ & $\vert \psi_{-,x} \,\rangle $\\   
$ ( -\pi/2)_{1y}$, $ ( \pi/2)_{2y}$  & $\vert \downarrow_x \uparrow_x \, \rangle $  & $\Delta = J/2$  & $\vert S_0 \,\rangle  $\\ 
$ ( \pi/2)_{1y}$, $ (-\pi/2)_{2y}$  & $\vert \uparrow_x \downarrow_x \, \rangle $  & $\Delta = J/2$  & $\vert T_{0,x} \,\rangle $\\  
$ ( \pi/2)_{1y}$, $ ( \pi/2)_{2y}$  & $\vert \uparrow_x \uparrow_x \, \rangle $  & $\Sigma = J/2$  & $\vert \psi_{+,x} \,\rangle  $\\  
\hline 
\end{tabular}
\caption{ 
Implementation of Step 3 in Fig.~\ref{fig2} for different resulting Bell states. The third column refers to the parameters of the detuned Hartman-Hahn pulse (DHH) introduced in Section~\ref{theorynon}.
} 
\label{table1}
\end{table}

\subsection{Quantum tomography of Bell states}
\label{tomography}

\begin{figure}[h!]
    \centering
    \begin{tikzpicture}  
    \node[inner sep=0pt] (duck) at (0,0)
    {\includegraphics[width=0.85\columnwidth]{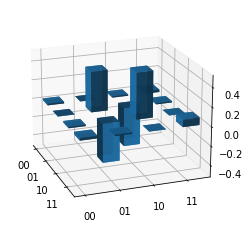}};
    \node[align=center,fill=white] at (-3.8, 1.8) {\textbf{(a)}};
    \end{tikzpicture}

    \begin{tikzpicture}
    \label{fig4b}
    \node[inner sep=0pt] (duck) at (0,0)
    {\includegraphics[width=0.85\columnwidth]{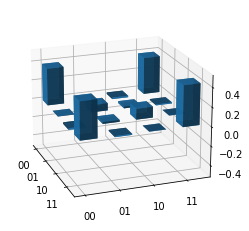}};
    \node[align=center,fill=white] at (-3.8, 1.8) {\textbf{(b)}};
    \end{tikzpicture}

    \caption{
\label{fig4.1}
  Real parts of the normalized density matrix of pseudo-pure states (a) $\vert S_0  \, \rangle $   and (b) $\vert \psi_{+,z}  \, \rangle $  obtained by quantum tomography for the two-spin system $^1$H~--$^{\ 13}$C. Labels $0$ and $1$ denote respectively the states $|\uparrow_z\rangle$ and $|\downarrow_z\rangle$.
}
\end{figure}

To verify experimentally that the prepared states are indeed the entangled Bell states, we performed on them the full quantum-state tomography according to the methodology of Ref.~\cite{PhysRevA.69.052302}. The experimentally measured tomograms of the real parts for the density matrices $\vert S_0  \, \rangle \langle  S_0 |$ and  $\vert \psi_{+,z} \, \rangle \langle \psi_{+,z} | $   in the ``local" basis 
$ \{ 
| \! \uparrow_z \uparrow_z \, \rangle$, 
$|\! \uparrow_z \downarrow_z \, \rangle$, 
$|\! \downarrow_z \uparrow_z \, \rangle$, 
$|\! \downarrow_z \downarrow_z \,\rangle  
\}$ are shown in Fig.~\ref{fig4.1}. The measured tomograms are in satisfactory agreement with the theoretical expectations, in particular,  as far as the entanglement-related off-diagonal elements are concerned.

The deviations of the measured density matrices exhibited in Fig.~\ref{fig4.1} from the idealized theoretical expressions indicate of the overall error of roughly 10 percent in implementing the pulse sequence (\ref{17}). There are two main reasons for this error: (i) the spatial inhomogeneity of the applied rf field, and (2) instrumental limitations constraining the control of the rf field amplitude to a discrete set of values. Both have affected the accuracy of the steps involving the resonant and the detuned Hartmann-Hahn conditions.

\section{Relaxation of Bell states: experiment and discussion} 
\label{relaxation}

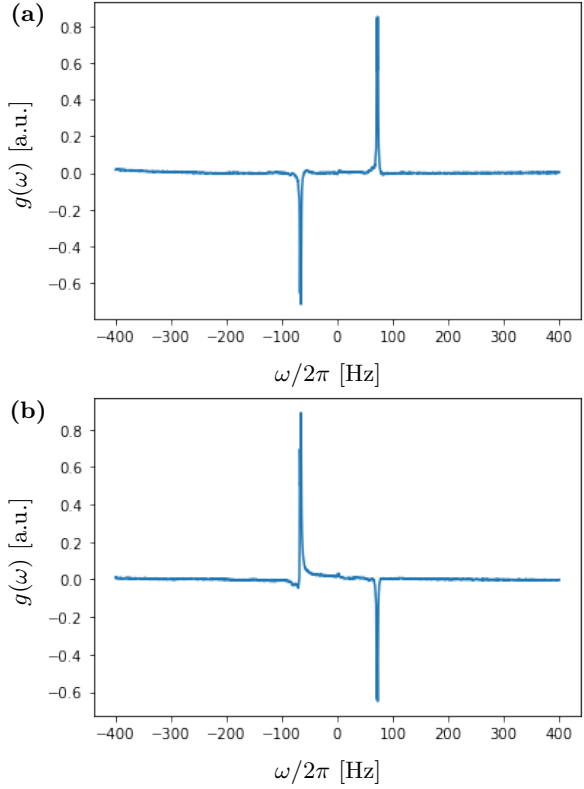
\begin{figure}
    \centering
    \begin{tikzpicture}  
    \node[inner sep=0pt] (duck) at (0,0)
    {\includegraphics[width=0.85\columnwidth]{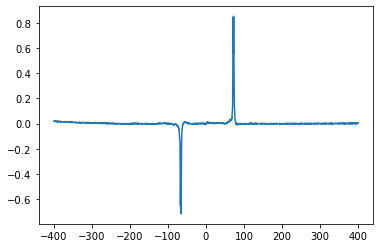}};
    \node[align=center,fill=white] at (-3.8, 2.1) {\textbf{(a)}};
    \node[align=center,fill=white] at (-3.9, 0.3) {{\rotatebox {90}{\text{$g(\omega)$ [a.u.]} }}};
    \end{tikzpicture}
\text{\qquad \quad $\omega/2\pi$ [Hz] }

    \begin{tikzpicture}
    \label{3b}
    \node[inner sep=0pt] (duck) at (0,0)
    {\includegraphics[width=0.85\columnwidth]{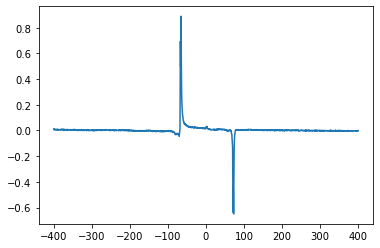}};
    \node[align=center,fill=white] at (-3.8, 2.1) {\textbf{(b)}};
    \node[align=center,fill=white] at (-3.9, 0.3) {{\rotatebox {90}{\text{$g(\omega)$ [a.u.]} }}};
    \end{tikzpicture}
\text{\qquad \quad $\omega/2\pi$ [Hz] }
    \caption{
\label{fig3}
Spectra $g(\omega)$ of the measured $^1$H as functions of the frequency offset $\omega$ for the following pseudo pure Bell states:  (a) $\vert S_0 \, \rangle$ and (b)  $\vert \psi_{+,z} \, \rangle$.
}
\end{figure}

\begin{figure} 
    \centering
    \begin{tikzpicture}
    \node[inner sep=0pt] (duck) at (0,0)
    {\includegraphics[width=0.85\columnwidth]{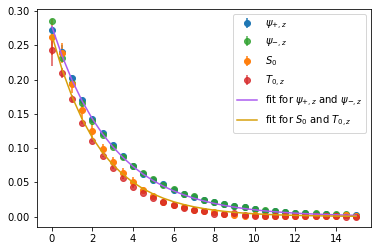}};
    \node[align=center,fill=white] at (-3.95, 2.1) {\textbf{(a)}}; 
    \node[align=center,fill=white] at (-3.9, 0.3) {{\rotatebox {90}{\text{$G_a$ [a.u.]} }}};
    \end{tikzpicture}
    \text{\qquad \quad $\tau$ [s] }
    \begin{tikzpicture}
    \node[inner sep=0pt] (duck) at (0,0)
    {\includegraphics[width=0.85\columnwidth]{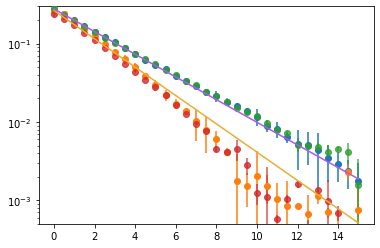}};
    \node[align=center,fill=white] at (-3.8, 2.1) {\textbf{(b)}};
    \node[align=center,fill=white] at (-3.9, 0.3) {{\rotatebox {90}{\text{$ G_a $ [a.u.]} }}};
    \end{tikzpicture}
    \text{\qquad \quad $\tau$ [s] }
    \caption{
\label{fig4}
Relaxation of the antisymmetric normalized component $G_{\text{a}}(\tau)$ of the $^1$H spectra for  Bell states $\vert S_0 \, \rangle$, $\vert T_{0,z} \, \rangle$, $\vert \psi_{+,z} \, \rangle$ and $\vert \psi_{-,z} \, \rangle$: (a) linear plots,  (b) semilogarithmic plots of the same data. Solid lines are the exponential fits to the initial 6 seconds used to extract the initial relaxation times. 
}
\end{figure}
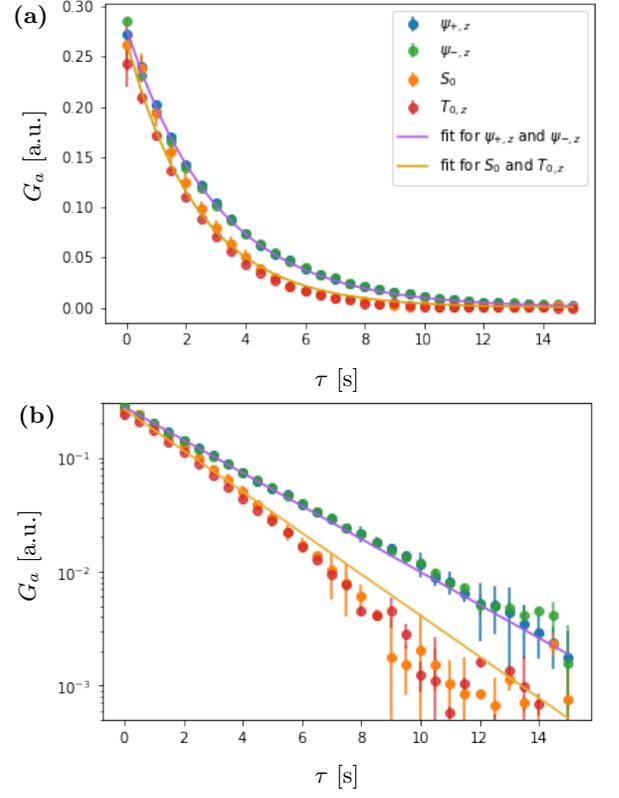

To monitor the relaxation of the Bell states, we did not measure the full tomograms as a function of time but rather recorded the spectra $g(\omega)$ of the free induction decay after the $ (\pi/2)_{1y}$ pulse on $^1$H. The $^1$H spectra of all four Bell states have antisymmetric character: for the states   $\vert S_0 \, \rangle$ and $\vert T_{0,z} \, \rangle$ the left peak (lower frequency) is negative, while the right peak (higher frequency) is of the same intensity but positive; the asymmetry is the opposite for the states $\vert \psi_{+,z} \, \rangle$ and $\vert \psi_{-,z} \, \rangle$. The experimental spectra $g(\omega)$ for the states $\vert \psi_{+,z} \, \rangle$ and $\vert S_0 \, \rangle$ right after the preparation sequence are shown in Fig.~\ref{fig3}.  When measured after a time delay $\tau$, the $^1$H spectra acquire a symmetric component as a result of the ordinary \mbox{$T_1$-relaxation} of $^{13}$C, while their antisymmetric components decay. We use the decay of the antisymmetric component of the spectrum 
\begin{equation}
  G_{\text{a}} \equiv \left| \ 
  \int_0^{+\infty} g(\omega) d\omega \ \ - \  \int_{-\infty}^0 g(\omega) d\omega \ 
  \right|
    \label{spectrum}
\end{equation}
as an indicator for the relaxation of the initially generated Bell state. (The numerical integration  was carried out from 20~Hz to 100~Hz for the first integral and from -100~Hz to -20~Hz for the second one in terms of the frequency offset used in Fig.~\ref{fig3}.)

The experimentally measured decays of the antisymmetric component of the $^1$H spectra  $G_{\text{a}}(\tau)$ for all four Bell  states are presented in Fig. \ref{fig4}. There one can see that the measurements for  $\vert T_{0,z} \, \rangle$  closely follow those for $\vert S_0 \, \rangle$, while the measurements for $\vert \psi_{-,z} \, \rangle$  agree with the ones for $\vert \psi_{+,z} \, \rangle$. To quantify the differences between the two pairs of datasets, we extracted the  {\it initial} relaxation times of $G_{\text{a}}(\tau)$ from the best exponential fits to the initial 6 seconds of decay. The extracted values  are $\tau_{S_0} = 2.4$~s from the combined datasets for $\vert S_0 \, \rangle$ and $\vert T_{0,z} \, \rangle$,  and $\tau_{\psi_+} = 3.0$~s from the combined datasets for $\vert \psi_{+,z} \, \rangle$ and $\vert \psi_{-,z} \, \rangle$.

In order to interpret the above relaxation measurements, we note that the density matrices of the four Bell states can be presented in the operator form as:
  \begin{equation}
      \label{C18}
      \begin{split}
                \vert S_0 \, \rangle \langle \, S_0 \vert =
          \frac{1}{4} \mathbb{I} - S_{1z}S_{2z} - \frac{1}{2}(S_{1+}S_{2-} + S_{1-}S_{2+}), \, \\
          \vert T_{0,z} \, \rangle \langle \, T_{0,z} \vert =
          \frac{1}{4} \mathbb{I} - S_{1z}S_{2z} + \frac{1}{2}(S_{1+}S_{2-} + S_{1-}S_{2+}), \,\\ 
          \vert \psi_{+,z} \, \rangle \langle \, \psi_{+,z} \vert =
          \frac{1}{4} \mathbb{I} + S_{1z}S_{2z} + \frac{1}{2}(S_{1+}S_{2+} + S_{1-}S_{2-}), \\
          \vert \psi_{-,z} \, \rangle \langle \, \psi_{-,z} \vert =
          \frac{1}{4} \mathbb{I} + S_{1z}S_{2z} - \frac{1}{2}(S_{1+}S_{2+} + S_{1-}S_{2-}), 
       \end{split}
  \end{equation}
where $S_{i \pm} \equiv S_{i x} \pm S_{i y}$.  The antisymmetric component of the measured $^1$H spectra $G_{\text{a}}$ originated in from the term $S_{1z}S_{2z}$ in each of the Eqs.(\ref{C18}). In other words, our measurements only characterized the relaxation of two-spin correlations in the diagonal part of the density matrix in the $S_z$ quantization basis. The values of $G_{\text{a}}$ were initially the same for all four Bell states, because $G_{\text{a}} \propto |S_{1z}S_{2z}|$. From such a perspective, one could have expected that the relaxation times $\tau_{S_0}$ and $\tau_{\psi_+}$ would be the same, because they would be tied to the relaxation of the average value $\langle S_{1z} S_{2z} \rangle (t)$ in the same environment. What is then the origin of the measured difference between $\tau_{S_0}$ and $\tau_{\psi_+}$?

We believe this difference is consistent with the so-called ``cross-correlation"~\cite{KUMAR2000191,GHOSH2005125} between two relaxation mechanisms -- the one due to the direct magnetic dipole coupling within the $^1$H~--$^{\ 13}$C pair and another one due to external local fields affecting differently the two spins. The latter mechanism can involve fluctuating chemical shift anisotropy (likely on $^{13}$C) and/or coupling to other nuclear spins on the same molecule and/or spin-rotation relaxation. The detailed investigation of the cross-correlation effects presented in Refs.~\cite{KUMAR2000191,GHOSH2005125} implies the following possible explanation of our observations.

The relaxation of the four diagonal terms of the spin-pair density matrix towards the thermal equilibrium is not independent of each other.  These four terms are determined, in addition to the overall normalization condition, by the average values $\langle S_{1z} \rangle (t)$, $\langle S_{2z} \rangle (t)$ and $\langle S_{1z} S_{2z} \rangle (t)$ -- the latter being monitored in our experiment. These three averages are coupled by the following general set of the relaxation equations \cite{GHOSH2005125}: 
\begin{widetext}
 \begin{equation}
     \label{C17}
     \frac{d}{dt}\begin{pmatrix} \langle S_{1z} \rangle(t) \\ \langle S_{2z} \rangle(t) \\ \langle S_{1z}S_{2z} \rangle(t) \end{pmatrix} = -
\begin{pmatrix} \mu_1 & \sigma_{12} & \delta_1 \\ \sigma_{12} & \mu_2 &\delta_2 \\ \delta_1 & \delta_2 & \mu_{12} \end{pmatrix}
\begin{pmatrix} \langle S_{1z} \rangle(t) - \langle S_{1z} \rangle_{\text{eq}} \\ \langle S_{2z} \rangle(t) - \langle S_{2z} \rangle_{\text{eq}} \\ \langle S_{1z}S_{2z} \rangle(t) \end{pmatrix},
  \end{equation}
\end{widetext}
where $\langle S_{1z} \rangle_{\text{eq}} = \varepsilon_1$ and $\langle S_{2z} \rangle_{\text{eq}} = \varepsilon_2$ are the respective equilibrium spin polarisations,   $\mu_1$ and $\mu_2$ are the diagonal relaxation rates for $\langle S_{1z} \rangle$ and $\langle S_{2z} \rangle$, $\mu_{12}$ is the diagonal relaxation rate for $\langle S_{1z}S_{2z} \rangle $, $\sigma_{12}$ is the off-diagonal rate for the cross-relaxation between $\langle S_{1z} \rangle$ and $\langle S_{2z} \rangle$  and, finally, $\delta_{1}$ and $\delta_{2}$ are the cross-correlation rates coupling $\langle S_{1z}S_{2z} \rangle $ with $\langle S_{1z} \rangle$ and $\langle S_{2z} \rangle$ respectively. These rates originate from the cross-correlations between different relaxation mechanisms~\cite{GHOSH2005125}.

When $\delta_{1}=0$ and $\delta_{2}=0$, the average $\langle S_{1z}S_{2z} \rangle $ is uncoupled from $\langle S_{1z} \rangle$ and $\langle S_{2z} \rangle$ and hence relaxes with the same rate $\mu_{12}$ independent of the initial conditions. In this case, both relaxation times $\tau_{S_0}$ and $\tau_{\psi_+}$ would be equal to $1/\mu_{12}$.  However, when $\delta_{1} \neq 0$ and/or $\delta_{2} \neq 0$, one needs to look for the set of eigenmodes of the system of equations (\ref{C17}) and then expand the initial ``vector'' $(\langle S_{1z} \rangle, \langle S_{2z} \rangle, \langle S_{1z}S_{2z} \rangle  )$ in terms of those exponential eigenmodes.  The initial conditions for the relaxation of  $\vert S_0  \, \rangle \langle  S_0 |$ and $\vert T_{0,z}  \, \rangle \langle  T_{0,z} |$  are  $\langle S_{1z} \rangle=0$, $\langle S_{2z} \rangle=0$ and $\langle S_{1z}S_{2z} \rangle = - \frac{\varepsilon_1 + \varepsilon_2}{8}$, while for 
$\vert \psi_{+,z} \, \rangle \langle \psi_{+,z} | $  and 
$\vert \psi_{-,z} \, \rangle \langle \psi_{+,z} | $ they are 
$(\langle S_{1z} \rangle=0$, $\langle S_{2z} \rangle=0$ and $\langle S_{1z}S_{2z} \rangle = \frac{\varepsilon_1 + \varepsilon_2}{8}$. 
Although the eigenmodes are the same in the two cases, the expansions of the initial conditions entail different contributions from different eigenmodes and hence different apparent initial relaxation rates given by the following general equation:
\begin{equation}
      \label{T0}
          \frac{1}{\tau_{\text{init}}}  = \mu_{12} 
          - \delta_1 \frac{\langle S_{1z} \rangle_{\text{eq}}}{\langle S_{1z}S_{2z} \rangle(0)} 
          - \delta_2 \frac{\langle S_{2z} \rangle_{\text{eq}}}{\langle S_{1z}S_{2z} \rangle(0)}
          .
  \end{equation}
This rate depends on the sign of $\langle S_{1z}S_{2z} \rangle(0)$, which is negative for $\vert S_0  \, \rangle \langle  S_0 |$ and $\vert T_{0,z}  \, \rangle \langle  T_{0,z} |$ and positive for $\vert \psi_{+,z} \, \rangle \langle \, \psi_{+,z} \vert$ and $\vert \psi_{-,z} \, \rangle \langle \, \psi_{-,z} \vert$. As a result,
\begin{eqnarray}
\frac{1}{\tau_{S_0}} - \frac{1}{\tau_{\psi_+}} \ &=& \  2 \ \ 
\frac{
\delta_1 \ \langle S_{1z} \rangle_{\text{eq}} + 
           \delta_2 \  \langle S_{2z} \rangle_{\text{eq}}
           }{|\langle S_{1z}S_{2z} \rangle(0)|} 
           \nonumber
           \\
 &=&  16 \ \  \frac{
\delta_1 \ \varepsilon_1 + 
           \delta_2 \  \varepsilon_2
           }{\varepsilon_1 + \varepsilon_2} ,
           \label{rate-difference}
\end{eqnarray}
while
\begin{equation}
\frac{1}{\tau_{S_0}} + \frac{1}{\tau_{\psi_+}} \ = \  2 \ \mu_{12} .
           \label{rate-sum}
\end{equation}

The theoretical picture based on Eq.(\ref{C17}) implies that, if the relaxation function $G_{\text{a}}(\tau)$ for different Bell states is accurately measured over sufficiently long times, it should always be decomposable into three exponential modes characterized by the same relaxation rates but different weights. It is, perhaps, somewhat of a coincidence that, within the time range accessible in the present experiment, the relaxation curves for $\vert \psi_{+,z} \, \rangle $ and $\vert \psi_{-,z} \, \rangle $ in Fig.~\ref{fig4}(b) appear to be monoexponential instead of exhibiting a multi-exponential behavior. On the other hand, the curves for  $\vert S_0  \, \rangle $ and $\vert T_{0,z}  \, \rangle $ do exhibit a multi-exponential character, as expected.

\ 
  
\section{Conclusions} 
\label{conclusions}
In this article, we introduced theoretically and implemented experimentally a method for generating pseudo-pure Bell states of a pair of coupled nonequivalent nuclear spins $1/2$ using detuned Hartmann-Hahn resonant condition. We then measured the relaxation of the diagonal elements of thus obtained Bell states in a liquid and found that the Bell states $\vert S_0 \rangle$ and $\vert T_{0,z}  \, \rangle$  belonging to the zero-quantum sector exhibit the initial relaxation rate which is different from the one for the states $\vert \psi_{+,z} \, \rangle$ and $\vert \psi_{-,z} \, \rangle$ belonging to the double-quantum sector. This difference was interpreted as originating from cross-correlations between distinct relaxation mechanisms.
Such an interpretation implies that the differences in the relaxations of Bell states can serve as a tool for discriminating the relaxation contributions from various mechanisms in liquid-state NMR, and, more generally, as an additional relaxation marker in the studies of complex molecules.

\medskip


\end{document}